\documentclass[aps,prb,12pt,endfloats*]{revtex4-1}

\usepackage{graphicx}
\usepackage{xfrac}
\usepackage{amsmath} 
\usepackage{textcomp}

\makeatletter
\newcommand{\smallsym}[2]{#1{\mathpalette\make@small@sym{#2}}}
\newcommand{\make@small@sym}[2]{%
  \vcenter{\hbox{$\m@th\downgrade@style#1#2$}}%
}
\newcommand{\downgrade@style}[1]{%
  \ifx#1\displaystyle\scriptstyle\else
    \ifx#1\textstyle\scriptstyle\else
      \scriptscriptstyle
  \fi\fi
}

\begin{document}

\title{Increased low-temperature damping in yttrium iron garnet thin films}

\author{C. L. Jermain}
\email{clj72@cornell.edu}
\affiliation{Cornell University, Ithaca, New York 14853, USA}
\author{S. V. Aradhya}
\affiliation{Cornell University, Ithaca, New York 14853, USA}
\author{J. T. Brangham}
\affiliation{Ohio State University, Columbus, Ohio 43210, USA}
\author{M. R. Page}
\affiliation{Ohio State University, Columbus, Ohio 43210, USA}
\author{N. D. Reynolds}
\affiliation{Cornell University, Ithaca, New York 14853, USA}
\author{P. C. Hammel}
\affiliation{Ohio State University, Columbus, Ohio 43210, USA}
\author{R. A. Buhrman}
\affiliation{Cornell University, Ithaca, New York 14853, USA}
\author{F. Y. Yang}
\affiliation{Ohio State University, Columbus, Ohio 43210, USA}
\author{D. C. Ralph}
\affiliation{Cornell University, Ithaca, New York 14853, USA}
\affiliation{Kavli Institute at Cornell for Nanoscale Science, Ithaca, New York, 14853, USA}

\date{\today}

\begin{abstract}
We report measurements of the frequency and temperature dependence of ferromagnetic resonance (FMR) for a 15-nm-thick yttrium iron garnet (YIG) film grown by off-axis sputtering. Although the FMR linewidth is narrow at room temperature (corresponding to a damping coefficient $\alpha$ = (9.0 $\pm$ 0.2) $\times 10^{-4}$), comparable to previous results for high-quality YIG films of similar thickness, the linewidth increases strongly at low temperatures, by a factor of almost 30. This increase cannot be explained as due to two-magnon scattering from defects at the sample interfaces. We argue that the increased low-temperature linewidth is due to impurity relaxation mechanisms that have been investigated previously in bulk YIG samples. We suggest that the low-temperature linewidth is a useful figure of merit to guide the optimization of thin-film growth protocols because it is a particularly sensitive indicator of impurities.
\end{abstract}


\maketitle 

Yttrium iron garnet (Y$_3$Fe$_5$O$_{12}$, YIG) thin films are of considerable interest for applications in spintronics and magnonics, since YIG can have one of the lowest damping coefficients of any magnetic material at room temperature.\cite{Sparks1964}  High-quality films of YIG and related garnets with thicknesses on the 10's of nm scale and below can be grown by pulsed-laser deposition (PLD),\cite{Kelly2013,Dorsey1993,Manuilov2009,Manuilov2010,Heinrich2011,Onbasli2014,Howe2015,Tang2016} off-axis sputtering,\cite{Wang2013,Wang2014,Liu2014,Chang2014,Brangham2016} and molecular-beam epitaxy (MBE).\cite{Jermain2016} Ferromagnetic resonance (FMR) measurements at room temperature for films grown by all of these techniques show that the FMR linewidth increases with decreasing film thickness, and the frequency dependence of the linewidth has a nonlinear functional form for very thin films.\cite{Kelly2013,Haidar2015,Jermain2016} This behavior has been attributed to two-magnon scattering at the film interfaces that becomes increasingly dominant as the film thickness decreases.\cite{Arias1999,Mills2003}

We are interested in extending the use of ultra-thin YIG films to cryogenic temperatures, for example so that we can use scannning SQUID microscopy\cite{Kirtley2016} to study the manipulation of YIG devices by spin-orbit torques.\cite{Miron2011,Liu2012} In the course of this work we have found that even apparently high-quality YIG films, which possess small FMR linewidths at room temperature, can have linewidths that increase dramatically with decreasing temperature. The 15 nm YIG film featured in this paper has a linewidth that increases by a factor of 28 as the temperature is lowered from room temperature to 25 K. The linewidth of this thin YIG film also shows an increasingly nonlinear frequency dependence as the temperature is lowered. We argue that these strong temperature dependencies cannot be explained by two-magnon scattering from the YIG interfaces.  Instead, we suggest that the increased linewidth at low temperature is due to magnetic damping associated with impurity mechanisms that have been studied previously in bulk YIG samples.\cite{Dillon1959,Spencer1961,Sparks1964,Seiden1964} 

\begin{figure}
\includegraphics[width=0.6\columnwidth]{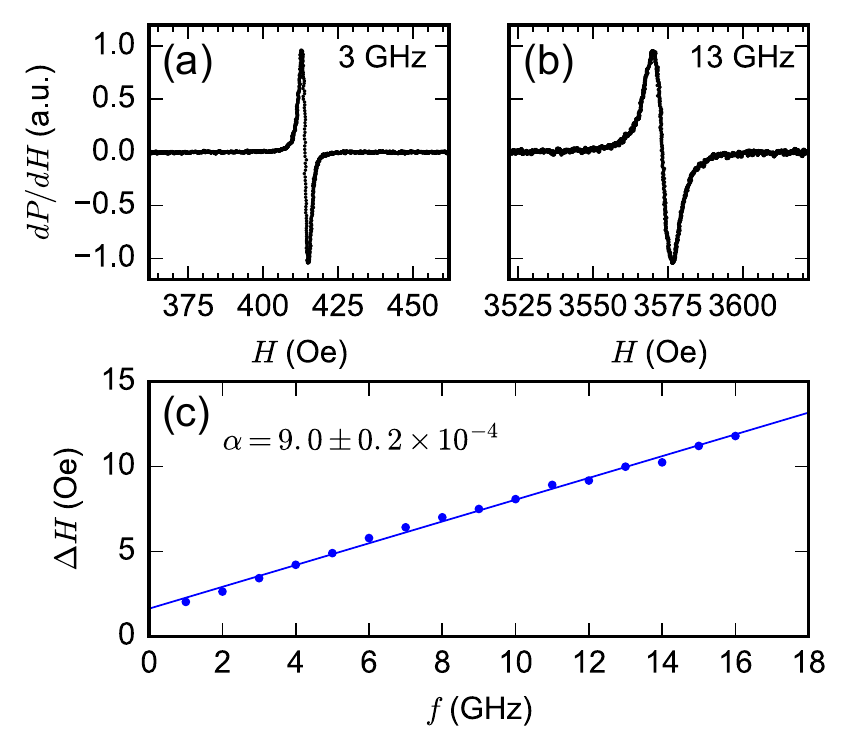}%
\caption{\label{figure-rt-fmr}Normalized ferromagnetic resonance spectra at (a) 3 GHz and (b) 13 GHz for a 15 nm YIG film at room temperature, with an in-plane applied magnetic field. (c) The frequency dependence of the linewidth corresponds to an effective Gilbert damping constant $\alpha$ = (9.0 $\pm$ 0.2) $\times 10^{-4}$.}%
\end{figure}

We grow our YIG films by off-axis sputtering\cite{Wang2013,Wang2014,Brangham2016} on a (111)-oriented gadolinium gallium garnet (GGG, Gd$_3$Ga$_5$O$_{12}$) substrate (see details in supplementary material). We measure the FMR response using a broadband coplanar waveguide with simultaneous field and power modulation.\cite{Jermain2016} The waveguide is installed in a continuous-flow He cryostat for temperature-dependent studies. Figure~\ref{figure-rt-fmr}(a) and (b) show room temperature FMR results at 3 and 13 GHz respectively, for a 15 nm film as deposited (\textit{i.e.}, without post-annealing). The resonances correspond well to derivatives of individual Lorentzians, to which we fit to extract the linewidth and resonance field. In Fig.~\ref{figure-rt-fmr}(c) we plot the Lorentzian full-width at half maximum (FWHM) linewidth $\Delta H$ versus frequency.  The slope of this curve corresponds to an effective Gilbert damping parameter $\alpha$ = (9.0 $\pm$ 0.2) $\times 10^{-4}$. This agrees well with previous measurements of a 14.0 nm YIG film grown by off-axis sputtering,\cite{Liu2014} which had a damping parameter $\alpha$ = (11.6 $\pm$ 0.7) $\times 10^{-4}$, and is within the range of measurements on PLD films of similar thickness.\cite{Kelly2013}

\begin{figure}
\includegraphics[width=0.6\columnwidth]{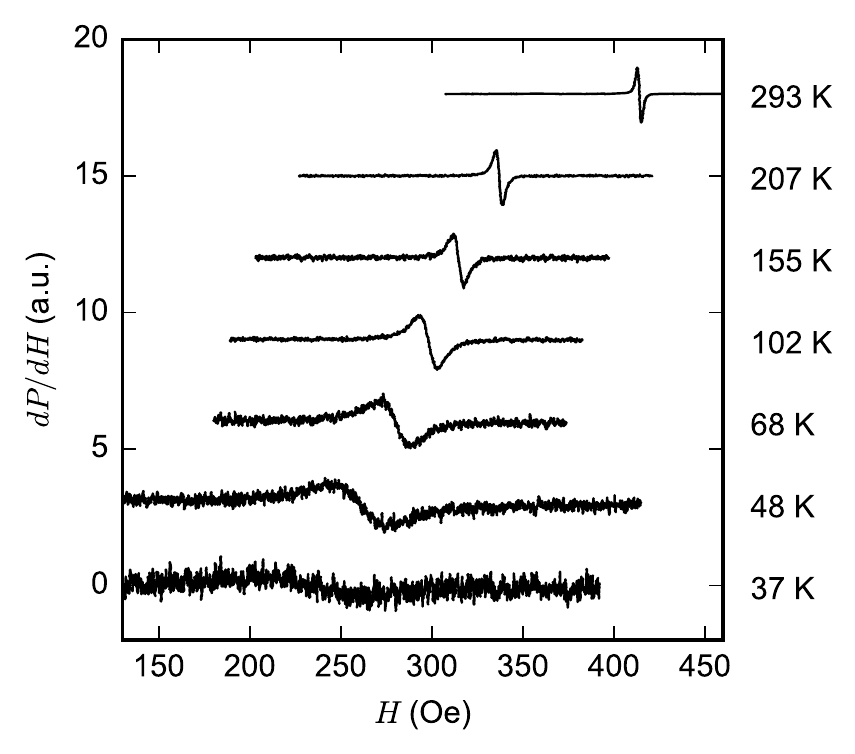}%
\caption{\label{figure-fmr}Normalized ferromagnetic resonance spectra at 3 GHz with an in-plane applied magnetic field for the YIG film at different temperatures. Different normalization factors are used for data at different temperatures; the actual amplitude of the resonances decreases strongly with decreasing temperature, as reflected in the decreasing signal-to-noise ratio. With decreasing temperature, we observe a large increase in the resonance linewidth.}%
\end{figure}

Figure~\ref{figure-fmr} shows how the in-plane FMR spectra of the same YIG film vary as a function of temperature. With decreasing temperature the data show a very large increase in the linewidth $\Delta H$, a shift in the resonance field, and a reduction in the amplitude of the signal, visible in the normalized curves as a reduction of the signal-to-noise ratio. The reduction in signal amplitude is consistent with the linewidth increase, given that the amplitude is expected\cite{Stancil2009} to scale with $(\Delta H)^{-2}$. Below roughly 37 K, the resonances become so broad that they are no longer distinguishable using the coplanar waveguide system. This strong temperature dependence is similar to results reported by Shigematsu \textit{et al.},\cite{Shigematsu2016} but it is not universal in ultra-thin YIG films: \textit{e.g.,} Haidar \textit{et al.}\ have observed in YIG films grown by PLD a damping coefficient that decreased by approximately a factor of two upon decreasing $T$ from room temperature to 8 K.\cite{Haidar2015}

\begin{figure}
\includegraphics[width=0.6\columnwidth]{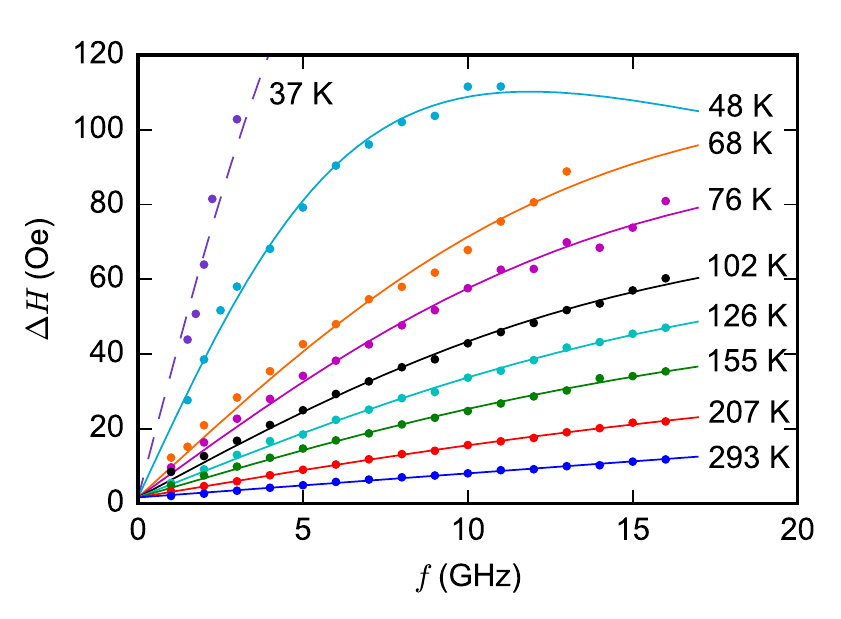}%
\caption{\label{figure-linewidths}Linewidths (Lorentzian FWHM) from the in-plane FMR spectra measured at different temperatures. Solid lines are fits to the sum of the frequency dependence expected from a slowly-relaxing impurity mechanism in addition to the room temperature linear behavior. The dashed line for 37 K is a guide to the eye.}%
\end{figure}

By analyzing similar FMR resonances obtained at different values of microwave frequency, we can extract both the frequency and temperature dependencies of $\Delta H$  (Fig.~\ref{figure-linewidths}). The frequency dependence at room temperature has an approximately linear dependence, similar to previous studies of high-quality YIG thin films in this thickness range.\cite{Sun2012,Kelly2013,Wang2014} As a function of decreasing temperature not only does the overall magnitude of the linewidth grow by a large factor, but at the same time there are strong deviations from linearity in the frequency dependence. These nonlinearities are qualitatively similar to what one might expect from two-magnon scattering from defects at the interfaces of the YIG film, but as we will argue below this mechanism cannot explain the very strong variations with temperature. We will instead argue that these changes can be accounted for by impurity relaxation within the YIG film.

We can obtain greater sensitivity in the FMR experiments, and thereby extend our study to temperatures lower than 37~K, by performing measurements in an X-band cavity. This comes at the cost of operating at fixed frequency (9.4 GHz). We perform background subtraction using in- and out-of-plane measurements in the cavity, as described in the supplementary material. Figure~\ref{figure-cavity} shows the $T$ dependence of the FMR linewidth in these cavity measurements, with a comparison to the broadband coplanar waveguide results. (The waveguide values are interpolated from measurements at 9 and 10 GHz.)  We find excellent quantitative agreement between the two types of measurements. The cavity measurements reveal that $\Delta H$ has a  maximum near 25~K, with a clear decrease at lower temperatures. The maximum linewidth is 28 times larger than the room temperature result at this frequency.

\begin{figure}
\includegraphics[width=0.6\columnwidth]{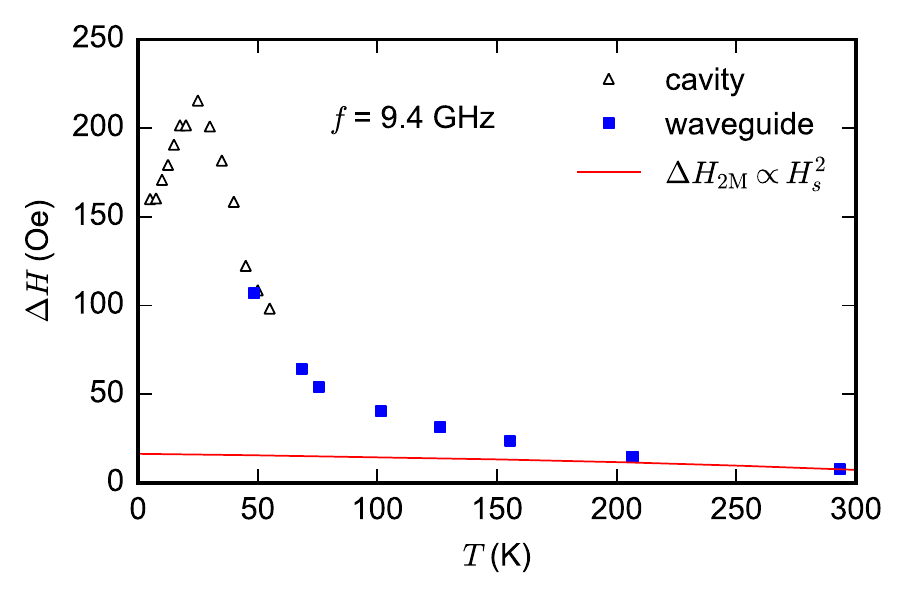}%
\caption{\label{figure-cavity}FMR linewidth at 9.4 GHz as measured by two techniques: (open black triangles) cavity measurements and (blue squares) coplanar waveguide measurements. We observe a peak near 25 K, where $\Delta H$ is  28 times larger than at room temperature. The solid red line indicates temperature dependence expected from two-magnon scattering; this dependence is too weak to explain the variation in $\Delta H$.}%
\end{figure}

In order to evaluate possible mechanisms for these very strong changes in linewidth with temperature, we must first characterize how the magnetic anisotropy in the YIG film varies with temperature.  We do this based on the  measured FMR resonance fields, fitting to the Kittel equation for a magnetic thin film with an in-plane magnetic field\cite{Stancil2009}

\begin{equation}
\label{equ-kittel}
f = \frac{|\gamma|}{2\pi}\sqrt{H_r^\parallel (H_r^\parallel + 4\pi M_\mathrm{eff})}.
\end{equation}

\begin{figure}
\includegraphics[width=0.6\columnwidth]{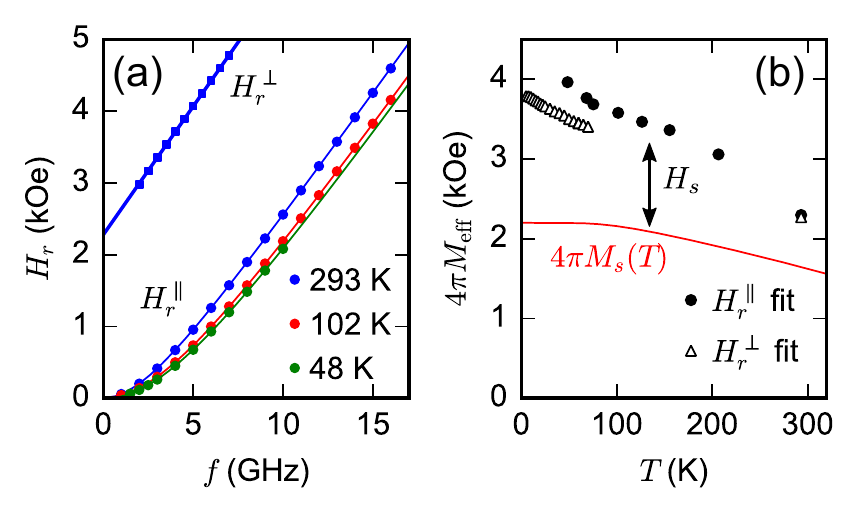}%
\caption{\label{figure-resonances}FMR resonance field as a function of frequency for (squares) an out-of-plane applied magnetic field at room temperature and (circles) and in-plane applied fields at various temperatures. Solid lines are fits to the Kittel equation. (b) Temperature dependence of the effective magnetization, determined from the Kittel fits for (black circles) in-plane and (open triangles) out-of-plane applied magnetic fields. The open triangles below 50 K are from cavity measurements. The red line is $4\pi$ times the saturation magnetization, from a fit to VSM measurements (see the supplementary material). The effective magnetization reflected in the magnetic anisotropy is significantly greater than the saturation magnetization.}%
\end{figure}

\noindent Here $\gamma$ is the gyromagnetic ratio, $H_r^\parallel$ is the in-plane resonance field for a given fixed frequency $f$, and $4\pi M_\mathrm{eff}$ parameterizes the shape anisotropy and any additional contributions to the perpendicular magnetic anisotropy. We obtain good fits (see Fig.~\ref{figure-resonances}(a)) with no additional in-plane anisotropy contribution. In Eq.~(\ref{equ-kittel}) we do not include a renormalization shift in the resonance frequency that can result from two-magnon scattering because this is small on the scale important to our analysis.\cite{Arias1999,Mills2003}  We also neglect a small shift in resonance field  that can arise from a static dipole interaction between the YIG and the paramagnetic GGG substrate\cite{Marysko1989,Marysko1991} because this is also small, less than a 1\% shift for temperatures above 15 K (see the supplementary material). The values of $4\pi M_\mathrm{eff}$ we obtain from the fits to Eq.~(\ref{equ-kittel}) at different temperatures are shown in Fig.~\ref{figure-resonances}(b). We find that $4\pi M_\mathrm{eff}$ is significantly larger than the simple shape anisotropy generated by the YIG saturation magnetization, $4\pi M_s$ (determined from vibrating sample magnetometry (VSM) measurements presented in the supplementary material), indicating the presence of a positive uniaxial anisotropy, $H_s = 4\pi M_\mathrm{eff} - 4\pi M_s$, favoring an in-plane magnetization.  We have confirmed the value of $H_s$ and the form of its temperature dependence using FMR measurements with an out-of-plane magnetic field. Figure~\ref{figure-resonances}(a) shows the frequency dependence of the resonance position with an out-of-plane field at room temperature, and Fig.~\ref{figure-resonances}(b) shows the extracted value of $4\pi M_\mathrm{eff}$ as a function of temperature from both waveguide and cavity FMR measurements.  The large value of $H_s$ is greater than expected from surface anisotropy\cite{Jermain2016} or magneto-crystalline anisotropy\cite{Stancil2009} of cubic YIG alone, so we tentatively ascribe the result to a growth-induced anisotropy, such as caused by tetragonal distortion. This is consistent with predictions and observations in YIG films grown by PLD,\cite{Manuilov2009} where the anisotropy is highly dependent on the growth conditions. The temperature dependence of $H_s$ that we obtain is qualitatively consistent with the spin fluctuation model,\cite{Callen1966,Pappas1996} which  predicts $H_s(T) \propto [M_s(T)]^2$.

Given this characterization of $H_s(T)$, we can now evaluate whether two-magnon scattering from surface defects, a mechanism that is expected to be active for ultra-thin YIG films at room temperature,\cite{Kelly2013,Jermain2016} is capable of explaining the large increase in the linewidth $\Delta H$ that we observe at low temperature.  This effect causes a linewidth that is nonlinear with frequency $f$, following the form\cite{Lenz2006}

\begin{equation}
\Delta H_\mathrm{2M} = \Gamma(T) \sin^{-1}{\sqrt{
	\frac{\sqrt{\omega^2 + (\omega_0/2)^2} - \omega_0/2}
    	 {\sqrt{\omega^2 + (\omega_0/2)^2} + \omega_0/2}}},
\end{equation}

\noindent where $\omega = 2\pi f$ and $\omega_0 = \gamma 4\pi M_\mathrm{eff}(T)$.  The temperature dependence in this equation is dominated by the scattering coefficient $\Gamma(T)$, whose expected temperature dependence\cite{Arias1999} is  $\Gamma(T) \propto [H_s(T)]^2$. Given our determination of $H_s(T)$ above (using $M_\mathrm{eff}$ from $H_r^\parallel$ fits as a worst-case scenario), the temperature dependence expected from the two-magnon scattering mechanism is illustrated by the red line in Fig.~\ref{figure-cavity}.  This mechanism can explain at most a factor of 4 increase in the linewidth as the temperature is reduced from 300 to 0 K, far less than the factor of 28 that we observe.  It also is incapable of explaining the peak in $\Delta H$ we measure near 25 K.  Similar conclusions follow if one assumes\cite{Callen1966,Pappas1996} that $H_s(T) \propto [M_s(T)]^2$, together with our VSM measurements of $M_s(T)$. 

An alternative mechanism that can account for a much stronger temperature dependence for $\Delta H$ is impurity relaxation, for example due to rare earth or Fe$^{2+}$ impurities in the YIG film. Researchers in the 1960's produced a rich body of literature which shows that the linewidth in bulk YIG samples can increase dramatically at low temperatures when impurity relaxation is active.\cite{Dillon1959,Spencer1961,Kittel1960,Sparks1964,Seiden1964}  The frequency and temperature dependence of $\Delta H$ in our samples can be explained well using a model of slowly-relaxing impurities.\cite{Clogston1955,Sparks1964,Seiden1964} The contribution to the linewidth from this mechanism is expected to have the form

\begin{equation}
\label{equ-slow}
\Delta H_\mathrm{SR} = A(T) \frac{\omega \tau}{1 + (\omega \tau)^2},
\end{equation}

\noindent where $A(T)$ is a frequency-independent prefactor and $\tau$ is a temperature-dependent time constant.  The lines in Fig.~\ref{figure-linewidths} are fits assuming that the linewidths are governed by this functional form plus a linear-in-frequency temperature-independent background contribution equal to the room-temperature dependence (Fig.~\ref{figure-rt-fmr}(c)).  The fit parameters are shown in the supplemental material. The maximum near 25 K in the temperature dependence of $\Delta H$ (Fig.~\ref{figure-cavity}) is very similar to previous measurements in bulk YIG,\cite{Sparks1964} and corresponds within the slowly-relaxing impurity model to the condition $\omega \tau \approx 1$.

In conclusion, even when a YIG film has a narrow linewidth at room temperaure indicating an apparently high-quality film, the linewidth can still increase dramatically at low temperature, by well over an order of magnitude. This is generally undesirable. For example, this will make manipulation of YIG films by anti-damping spin-transfer torques much less efficient at low temperature, and may block it entirely for practical purposes. Based on measurements of the temperature and frequency dependence of the effect, we suggest that the increased low-temperature linewidth is due to slowly relaxing impurities, perhaps rare earth or Fe$^{2+}$ impurities introduced during growth.\cite{Sparks1964} Given the high degree of sensitivity of the low-temperature linewidth to these impurities, we suggest that the low-$T$ linewidth can serve as a useful figure of merit for optimizing growth protocols for ultra-thin YIG films.

\section*{Supplementary Material}

See the supplementary material for detailed information on the YIG growth by off-axis sputtering, saturation magnetization and substrate susceptibility, broadband and cavity measurement systems, resonance field shift from substrate dipolar fields,  thickness dependence of the linewidths, and fitting parameters extracted from the frequency dependence of the linewidths.

\begin{acknowledgments}
We acknowledge G. E. Rowlands and S. Shi for their help in building a low-temperature cryostat, M. S. Weathers for her help with X-ray reflectivity measurements, and B. Dzikovski for his help with the low-temperature cavity measurements preformed at ACERT (National Biomedical Center for Advanced ESR Technology). This research was supported by the National Science Foundation (DMR-1406333 and DMR-1507274), NSF/MRSEC program through the Center for Emergent Materials (DMR-1420451), the NSF/MRSEC program through the Cornell Center for Materials Research (DMR-1120296), the NIH/NIGMS program through ACERT (P41GM103521), the US Department of Energy (DOE) (DE-FG02-03ER46054), and the US Department of Defense - Intelligence Advanced Research Projects Activity (IARPA) through the U.S. Army Research Office (W911NF-14-C-0089). The content of the paper does not necessarily reflect the position or the policy of the Government, and no official endorsement should be inferred. This work made use of the Cornell Center for Materials Research shared facilities. 
\end{acknowledgments}

\bibliography{lt-yig.bib}

\end{document}